# The extended narrow line region in Mkn79.

## I.Observations.

**L.S.Nazarova,[1,2] P.T.O'Brien,[3] and M.J.Ward[3]**

Royal Greenwich Observatory, Madingley Rd.,Cambridge,CB3 0EZ,UK
Sternberg Astronomical Institute,Universitetskij prosp.13, Moscow,119899, Russia
Astrophysics, Department of Physics, University of Oxford, Keble Road, Oxford, OX1 3RH, UK

**Abstract.** We present deep long slit spectra of Mkn79 in position angles PA=12$^o$ and PA=50$^o$ obtained with the WHT. These data prove the existence of an extended narrow line region in PA=12$^o$, which coincides with the triplet radio structure (Ulvestad & Wilson 1984) and the observed outflow of material from the nucleus at PA=10$^o$ (Whittle et al. 1988). The ratios of the high to low ionization lines indicate a higher level of gas excitation in PA=12$^o$ compared to PA=50$^o$. The [NII]$\lambda$6583/H$\alpha$ and [SII]$\lambda$6717,31/H$\alpha$ versus [OIII]$\lambda$5007/H$\beta$ line ratios are consistent with excitation by an AGN continuum rather than a HII region.

### 1. Introduction.

Research on the emission spectra of Active Galactic Nuclei (AGN) has led to the assumption that the emission-line gas can be separated into three regions: the Broad Line Region (BLR); the Narrow Line Region (NLR); and the Extended Narrow Line Region (ENLR). These regions have a different prominence in Seyfert 1 and Seyfert 2 galaxies. In white light, the BLR, NLR and in some cases the ENLR have been observed in Seyfert 1 galaxies, but Seyfert 2 galaxies show strong evidence only for the NLR and ENLR. When studying AGN the problem is to decide whether they are intrinsically different or are merely different manifestations of the same phenomenon. In recent years a belief has grown that the observed differences can be accommodated within a "Unified Scheme" . According to this model the diversity between Seyfert 1 and Seyfert 2 galaxies is only due to the varying degree of obscuration and orientation (Antonucci & Miller 1985; Antonucci 1993).

The morphology of the NLR and ENLR indicates that the emission-line gas is aligned with the radio axis and is probably illuminated by an anisotropic beam of ionizing UV- radiation. Indeed, the near-UV continuum (Pogge & De Robertis 1993) and X-ray images of NGC1068 (Wilson et al. 1992) show that the circumnuclear emission has a cone-like geometry, and an elongated morphology in the direction of the radio jet (Haniff et al. 1988,1991; Ulvestad et al. 1987; Wilson & Ulvestad 1987; Pogge 1988,1989). Similar results with long slit spectroscopy and deep narrow band filter CCD images have been obtained

*Send offprint requests to*: L.S.Nazarova

al. 1988; Petitjean & Durret 1993). Generally speaking there are about a dozen objects which show an elongated morphology in both the radio and optical (Wilson & Tsvetanov 1994).

The Seyfert 1.2 galaxy Mkn79 is one of those having an elongated radio morphology (Ulvestad & Wilson 1984). The radio map shows one component 1 arcsec to the south (PA=182$^o$) and another 1.9 arcsec to the north (PA=12$^o$) of the nucleus. Inspection of the velocity field from [OIII]$\lambda$5007 profiles in PA=10$^o$ shows double-component peak velocities across the nucleus with shifts relative to the systemic velocity of +100 km s$^{-1}$ and −50 km s$^{-1}$ (Whittle et al. 1988). However the [OIII]5007 and H$\alpha$ maps do not indicate the spatial extent of the emission (Haniff, Wilson & Ward 1988). This result could be caused by uncertain deconvolution from the seeing ($\geq$1.5”) when there is a small separation between the components ($\geq$2”). The systemic velocity of Mkn79 is 6643 km s$^{-1}$ (Heckman, Balick & Sullivan 1978), giving a spatial scale of 630 pc arcsec$^{-1}$, for $H_o$=50 km s$^{-1}$ Mpc$^{-1}$ and $q_o$=0. The rotation velocity obtained from the HI profile width is 220 km s$^{-1}$ (Heckman, Balick & Sullivan 1978). The host galaxy of Mkn79 is extended along a line with PA=65$^o$ east of north and with an axes ratio $b/a$= 0.69 (Keel 1980).

The existence of the ENLR in Mkn79 has not been discovered before, although the triplet radio structure (Ulvestad & Wilson 1984) and possible outflow, in addition to the normal galaxy rotation (Whittle et al. 1988), indirectly indicate that an ENLR might exist. In order to study any possible ENLR in Mkn79 we observed the galaxy using the WHT telescope on La Palma in two position angles – along the radio structure in PA=12$^o$ and in PA=50$^o$ which is close to the global extended structure.

In this paper we describe the observational material as well as discuss the ionization properties of the ENLR in Mkn79. The observations and data reduction are described the discussion and conclusion are given in Sections 4 and Section 5 respectively.

## 2. Observations and reductions

A series of long-slit spectra of Mkn79 were obtained with the ISIS double spectrograph using the EEV2 and TEK1 CCD detectors on the 4.2m WHT telescope on La Palma during service time on the night of 15-16 February 1994. The galaxy was observed in two positional angles of 12$^o$ and 50$^o$ (Fig.1).

**Fig. 1.** The central region of Mkn79 in blue light, from Mazzarella & Boroson (1993). The radial slit positions in PA=12$^o$ and PA=50$^o$ are superimposed.

The individual exposure times were limited to 1800 seconds in order to avoid saturating the chip. Spectra were obtained covering two wavelengths ranges; from 3700Å to 5230Å (blue arm) and from 6110Å to 7460Å (red arm). The total exposure times were 6400s and 5400s in PA=12$^o$ and PA=50$^o$ respectively. The weather conditions were contains 331 × 1001 pixels. Each pixel corresponds to 0.33 arcsec in the spatial direction, and 1.45Å in the spectral direction. The observation log is presented in Table 1.

**Table 1.** Log of observations on 15-16 February 1994 with WHT-ISIS

| Detector | Airmass | PA | Range(Å) | Exposure(s) | Slit |
|---|---|---|---|---|---|
| EEV3 | 1.165 | 12 | 6110 − 7460 | 1000 | 1.5 |
| TEK1 | 1.165 | | 3700 − 5230 | 1000 | |
| EEV3 | 1.138 | | 6110 − 7460 | 1800 | |
| TEK1 | 1.138 | | 3700 − 5230 | 1800 | |
| EEV3 | 1.089 | | 6110 − 7460 | 1800 | |
| TEK1 | 1.089 | | 3700 − 7460 | 1800 | |
| EEV3 | 1.074 | | 6110 − 7460 | 1800 | |
| TEK1 | 1.074 | | 3700 − 7460 | 1800 | |
| EEV3 | 1.081 | 50 | 6110 − 7460 | 1800 | |
| TEK1 | 1.081 | | 3700 − 5230 | 1800 | |
| EEV3 | 1.105 | | 6110 − 7460 | 1800 | |
| TEK1 | 1.105 | | 3700 − 5230 | 1800 | |
| EEV3 | 1.142 | | 6110 − 7460 | 1800 | |
| TEK1 | 1.142 | | 3700 − 5230 | 1800 | |

**Table 2.** Extraction windows in the 2-D images. Width of windows and distance of the centre of the windows from the centre of Mkn79.

| Label | PA | Width (*arcsec*) | Distance (*arcsec*) | Distance (*kpc*) |
|---|---|---|---|---|
| Region 12 − 1 | 12 | 4 | 15 | 9.45 |
| Region 12 − 2 | | 4 | 11 | 6.93 |
| Region 12 − 3 | | 4 | 7 | 4.41 |
| Region 12 − 4 | | 4 | 3 | 1.89 |
| Nucleus | | 2 | 0.0 | 0.0 |
| Region 12 − 5 | | 4 | −3 | 1.89 |
| Region 12 − 6 | | 4 | −7 | 4.41 |
| Region 12 − 7 | | 4 | −11 | 6.93 |
| Region 12 − 8 | | 4 | −15 | 9.45 |
| Region 50 − 1 | 50 | 4 | 15 | 9.45 |
| Region 50 − 2 | | 4 | 11 | 6.93 |
| Region 50 − 3 | | 4 | 7 | 4.41 |
| Region 50 − 4 | | 4 | 3 | 1.89 |
| Region 50 − 5 | | 4 | −3 | 1.89 |
| Region 50 − 6 | | 4 | −7 | 4.41 |
| Region 50 − 7 | | 4 | −11 | 6.93 |
| Region 50 − 8 | | 4 | −15 | 9.45 |

ware package (Shortridge 1993). Differential extinction effects were removed. De-biasing, flat-fielding and wavelength calibration were performed in the usual manner. The standard star was G191B2B with fluxes taken from Oke (1974). Spectra were then extracted from the two-dimensional images using a variety of extraction windows located at different distances along the slit. The primary off-nucleus spectra discussed in this paper were extracted in each PA using 8 windows 12 pixels wide (4 arcsec) as tabulated in Table 2. The nuclear spectrum was extracted with a window of 2 arcsec (=1.26 kpc). The relative line intensity ratios (using H$\beta$=100) for all of these spectra are given in Tables 3, 4 and 5.

The emission line fluxes were obtained by fitting gaussian profiles using the Longslit spectral analysis software (Wilkins & Axon 1992). Most lines were well fitted by single skew gaussians. However, multi-component fits were used for the [SII]$\lambda\lambda$6717,6731 doublet, for the H$\alpha$+[NII]$\lambda\lambda$6548,6583 complex, and to isolate the broad from the narrow components in the H$\beta$, H$\gamma$ and H$\delta$ lines. The multi-component fits were generally good, although it was difficult in the nucleus to separate the [NII]$\lambda$6548 and [NII]$\lambda$6583 lines from H$\alpha$ due to the large intensity of the broad H$\alpha$ component. The errors on the measured fluxes were taken from the fitting software.

In order to check the accuracy of the flux measurements we have compared the measured ratios of lines emitted from the same upper level with their known values, which reflect the radiative transition probabilities. The ratios of [OIII]$\lambda$5007/[OIII]$\lambda$4959 and [NII]$\lambda$6583/[NII]$\lambda$6548 lines are plotted against position along the slit in Fig.2. Our measurements are very close to the theoretical values of 2.94 and 2.88, and do not suggest any systematic trends with position. The scatter is entirely consistent with the error bars excluding a few cases which are possibly due to the uncertainty in separation be-tween lines [OIII]$\lambda$4959 and [OIII]$\lambda$5007. Small systematic errors when comparing lines from the separate wavelength ranges cannot be ruled out. This is particular important for ratios such as [SII]$\lambda$6725/H$\beta$ and [NII]$\lambda$6583/H$\beta$.

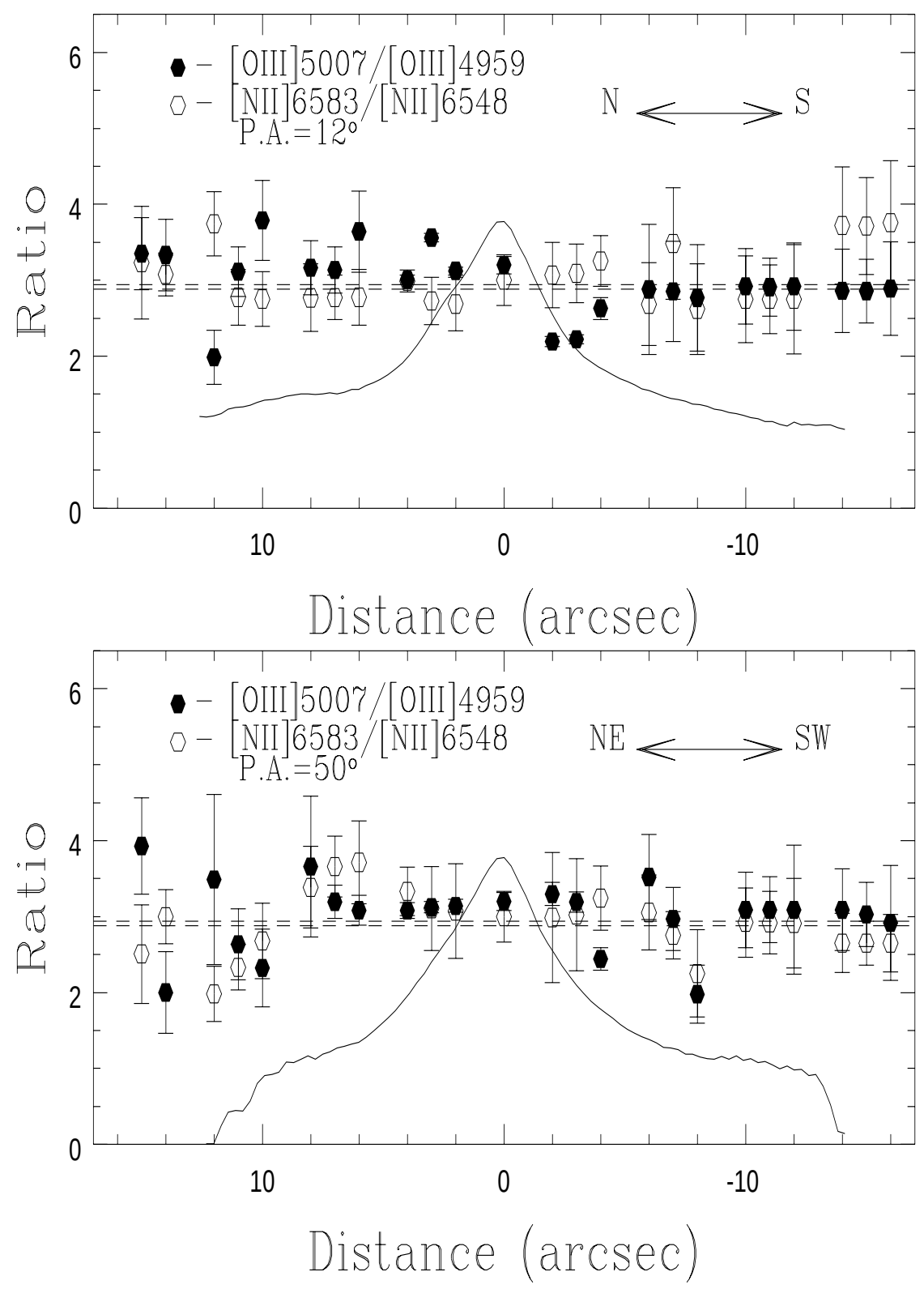

**Fig. 2.** Measured ratios of the [OIII]$\lambda$5007/[OIII]$\lambda$4959 and [NII]$\lambda$6583/[NII]$\lambda$6548 lines plotted against position along the slit. The theoretical values 2.94 and 2.88 are shown by dotted lines. The [OIII]$\lambda$5007 spatial flux profiles are shown by solid lines for both position angles.

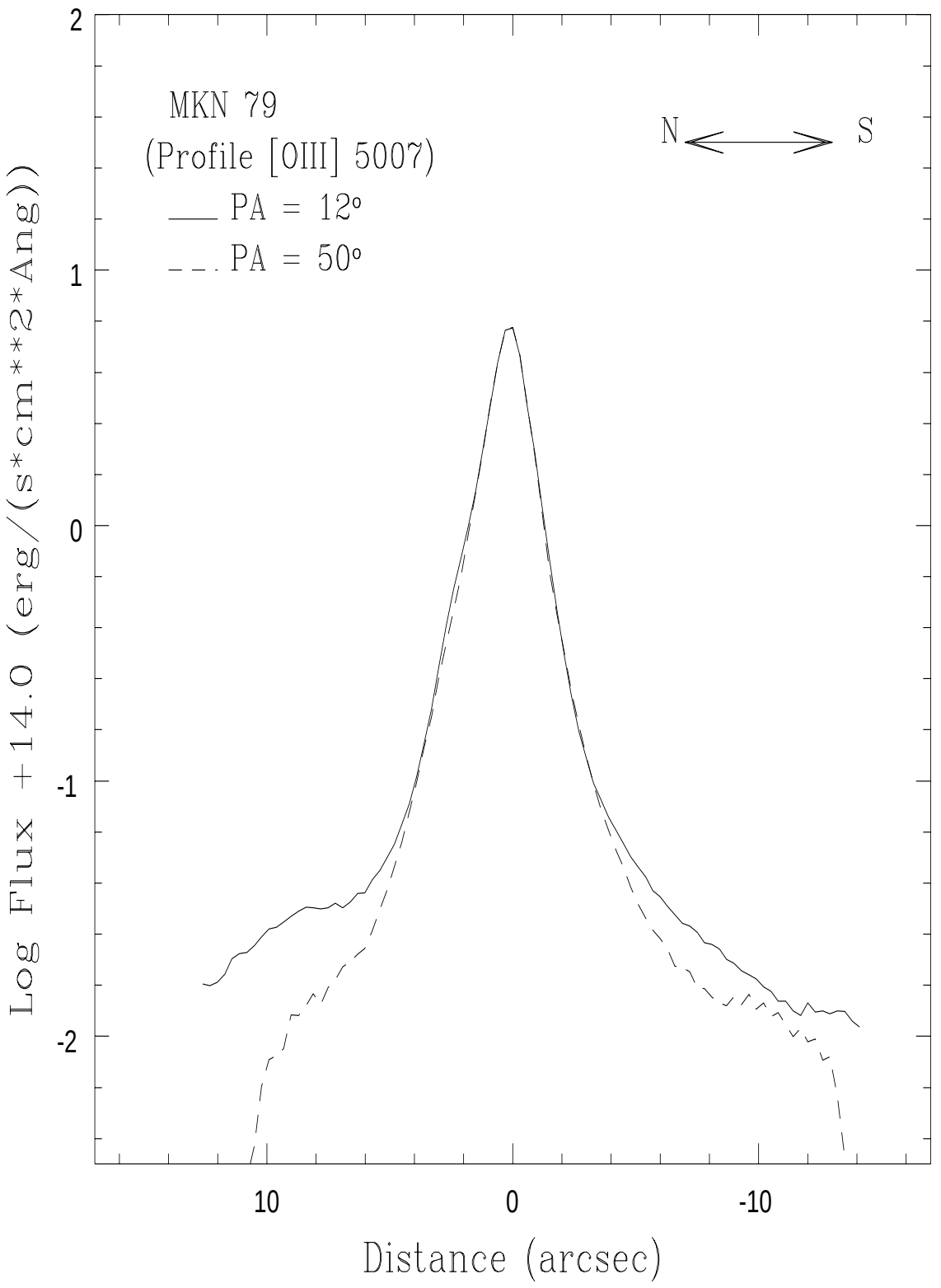

**Fig. 4.** The radial flux profiles of [OIII]$\lambda$5007 in the two position angles.

## 3. Results

This section presents the results of long-slit spectroscopy of Mkn 79 and the emission line measurements in two position angles. The line intensities have been obtained at different distances from the nucleus and in the nucleus itself.

### *3.1. 2-Dimensional Images*

The emission structure in the wavelength ranges around H$\beta$ and $H\alpha$ are shown in Fig.3 for both position angles. The comparison of the 2-D images in both position angles shows the existence of an extended structure in the [OIII]$\lambda\lambda$5007,4959 high ionization lines in PA=12$^o$ compared to PA=50$^o$. The extended structure in the [OIII]$\lambda$5007 line can be detected up to 11 arcsec or 6.8 kpc from the nucleus in PA=12$^o$, compared to approximately 7 arcsec in PA=50$^o$ for the same intensity contour. Seeing effects cannot be the reason for these differences as the observations were taken close together during a night with stable weather conditions. The radial flux profiles of the high ionization [OIII]$\lambda$5007 line shown in Fig.4 also indicate that the flux in PA=50$^o$ falls quicker with the distance from the centre than in PA=12$^o$.

In PA=50$^o$ there is also strong emission in the hydrogen lines, and in the low ionization [NII]$\lambda\lambda$6548,6583 and [SII]$\lambda\lambda$6717,6731 lines at some considerable distance from the centre. The line intensity ratios for these regions

**Fig. 3.** The emission structure of Mkn79 around H$\beta$ and H$\alpha$. North is at the bottom. a)PA=12$^o$ - blue range of the spectrum; b) PA=50$^o$ - blue range of the spectrum; c)PA=12$^o$ - red range of the spectrum; d)PA=50 - red range of the spectrum. Twenty equally spaced contours between 8$\times$10$^{-17}$erg cm$^{-2}$ Å$^{-1}$ and 8$\times$10$^{-18}$erg cm$^{-2}$ Å$^{-1}$ are plotted.

(Section 4), strongly suggesting an association with HII regions in the disk of the host galaxy. This PA coincides with that of the global extended structure of Mkn79 (the stellar disk), where HII regions should exist.

**Table 3.** Line intensities relative to I($H\beta_n$) and errors in the nucleus

| Ion | $\lambda^{(a)}$ | $I/I_{H\beta}$ |
|---|---|---|
| [OII] | 3813 | 164.5±16.3 |
| [NeIII] | 3956 | 71.3±6.7 |
| HeI | 3977 | 44.6±6.5 |
| [NeIII] | 4056 | 31.12±3.7 |
| H$\delta$(b) | 4177 | 330.5±30 |
| H$\delta$(n) | 4198 | 34.4±4.2 |
| H$\gamma$(b) | 4428 | 716.7±25 |
| H$\gamma$(n) | 4437 | 64.1±6.7 |
| [OIII] | 4459 | 40.5±4.1 |
| HeII | 4790 | 32.9±4.7 |
| H$\beta$(n) | 4970 | 100.0 |
| H$\beta$(b) | 4967 | 1961.2±186 |
| [OIII] | 5070 | 247.7±33.0 |
| [OIII] | 5121 | 792.7±73.0 |
| [FeVII] | 6221 | 25.48±2.6 |
| [OI] | 6440 | 54.86±4.9 |
| [OI] | 6505 | 17.9±1.0 |
| [FeX] | 6516 | 7.9±0.9 |
| [OI] | 6440 | 54.86±4.9 |
| [OI] | 6505 | 17.9±1.0 |
| [FeX] | 6516 | 7.9±0.9 |
| [NII] | 6695 | 67.97±7.9 |
| H$\alpha$(b) | 6695 | 6860.5±568.6 |
| H$\alpha$(n) | 6708 | 386.5±38.4 |
| [NII] | 6727 | 203.95±24.9 |
| [SII] | 6863 | 54.0±5.6 |
| [SII] | 6881 | 64.0±5.8 |
| [ArV] | 7159 | 51.1±4.9 |
| HeI | 7220 | 47.5±3.0 |
| [ArIII] | 7296 | 16.0±1.6 |
| F(H$\beta$) $^{(c)}$ | 4970 | 922.4±88.7 |

(a) Observed wavelength
(b) Broad Balmer line component
(n) Narrow Balmer line component
(c) Flux of narrow component of H$\beta$ in units of $10^{-17}$erg $s^{-1}$ $cm^{-2}$

*3.2. The Nucleus of Mkn79.*

The optical spectrum of Mkn79 has been studied by Oke & Lauer (1979). They found that the broad-line region in Mkn79 has an electron density ($N_e$) greater than $10^8$ $cm^{-3}$ in the ionized hydrogen zone and about $10^7$ $cm^{-3}$ in the neutral zone. They also found that the high-ionization narrow-line region has $N_e$= $10^5$ $cm^{-3}$ and its electron temperatures ($T_e$) in the range of 20,000-30,000K. However the low-ionization narrow-line region has electron temperature $T_e$=10,000K and electron density, $N_e$, from $10^3$ to $10^4$ $cm^{-3}$, derived from an analysis of the [OII],[NII] and [SII] line ratios.

The optical spectrum of the Mkn79 nucleus in the blue and red is presented in Fig.5. The nuclear spectrum is characterized by emission from high-ionization gas, including lines of HeII, [OIII], [Fe VII] and [Fe X]. A wide range of ionization stages are represented, such as $Fe^{+9}$ and $O^{+0}$. The nuclear line ratios are given in Table 3. We note that the Balmer broad to narrow line ratios differ from some others given in the literature (Oke & Lauer 1979; Cohen 1983; De Zotti & Gaskell 1985), probably due to broad line variability (Rosenblatt et al. 1992) and differences in spectral resolution, data quality and seeing which all affect spectral deconvolution. The broad component of H$\beta$ has an asymmetric profile with a shoulder in the red wing, and does not show a significant change in the asymmetry of this line (Oke & Lauer 1979). However the flux in the lines and in the continuum have significantly changed during the period 1979 - 1984 (Peterson et al. 1982, Peterson & Gaskell 1986, Rosenblatt et al. 1992). The nuclear continuum also clearly shows the 'small blue bump' probably due to FeII and Balmer continuum emission, as also found by Oke & Zimmerman (1979).

*3.3. The Extended Narrow Line Region in Mkn79.*

The emission spectra extracted at the 8 locations tabulated in Table 2 are shown in Fig. 6. The continuum emission and ionization level decrease with distance from the nucleus. Comparison of the spectra, for similar projected distances, in the two directions indicate that there is a significant difference in ionization level between them. In PA=12$^o$ the spectra have a stronger blue continuum than in PA=50$^o$. The line intensities are also higher in the north (regions 1 - 3) than in the south (regions 6 - 8 ). Overall, the ionization level is highest north of the nucleus in PA=12$^o$.

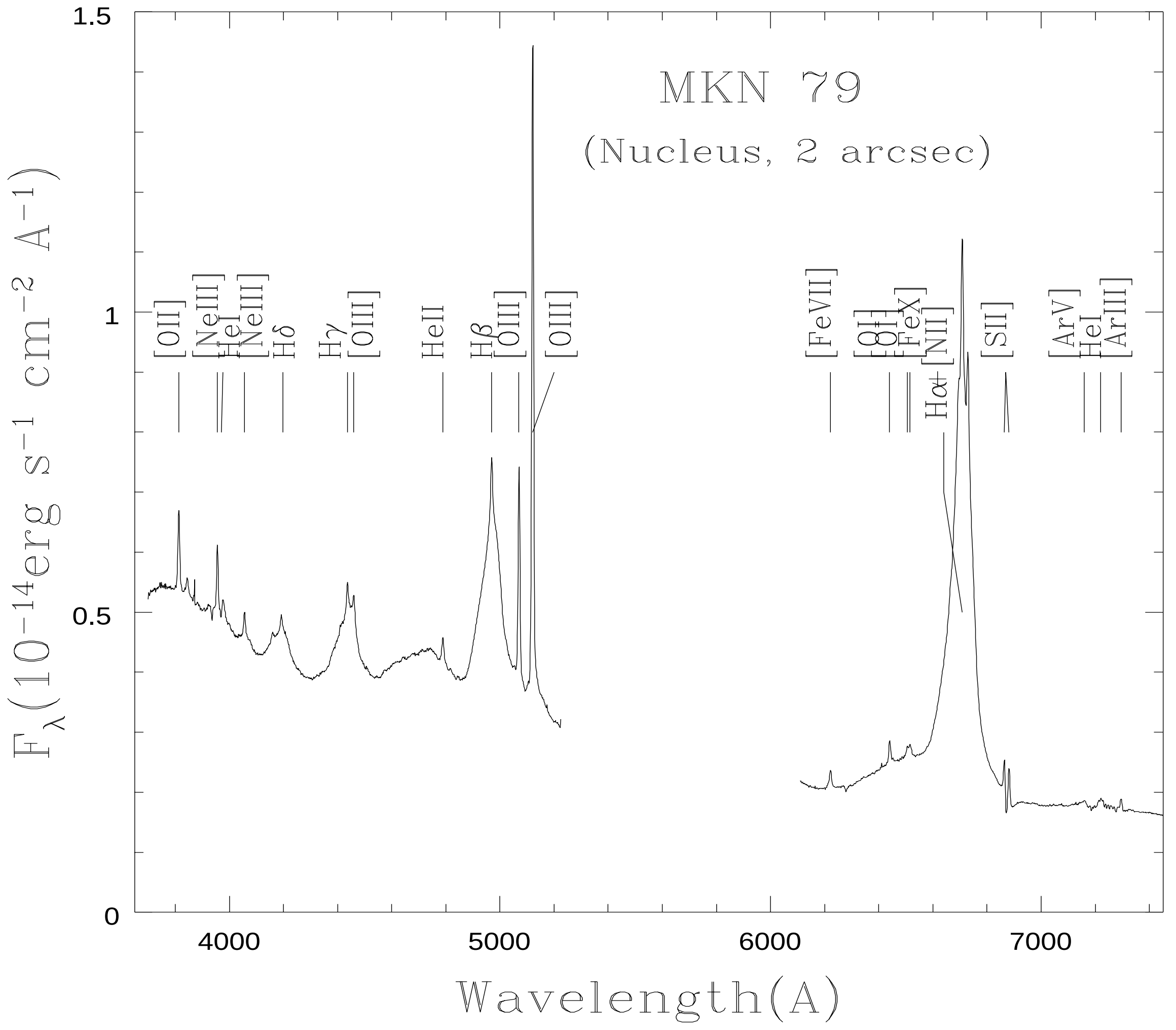

**Fig. 5.** The spectra of the nucleus of Mkn79. The extraction window corresponds to 2 arcsec.

The spectra in regions 4 and 5 are similar to the nucleus, and clearly show broad lines, presumably due to seeing effects. The differences between regions 4 and 5, and between PA=12$^o$ and PA=50$^o$ may be due to slight errors in slit placement, determination of the exact location of the nucleus, and differential refraction. With the exception of regions 4 and 5, the emission lines in all other regions are narrow and unresolved. Absorption lines of H & K Ca II are present in the spectra in both directions, but they are stronger in PA=50$^o$ where the relative host galaxy contribution could be higher.

Our results show that the ENLR Balmer decrement is nearly constant, within the errors, at different distances from the nucleus and in both directions. The average value of the Balmer decrement is about 3.1, corresponding to pure case B recombination. Therefore we did not make any corrections for reddening to the line fluxes, and associated ratios, presented in this paper.

Several emission line ratios are plotted as functions of cross-section in Figs.7 and 8. In Fig.7 we present the ratios

**Table 4.** Line intensities relative to I($H\beta_n$) and errors in PA=12$^o$

| Ion | $\lambda^{(a)}$ | Region 1 | Region 2 | Region 3 | Region 4 | Region 5 | Region 6 | Region 7 | Region 8 |
|---|---|---|---|---|---|---|---|---|---|
| [OII] | 3813 | 183±26 | 192±18 | 211±17 | 195±17 | 247±26 | 231±22 | 155±20 | 85±15 |
| [NeIII] | 3956 | | | | 68±3 | 124±13 | | | |
| HeI | 3977 | | | | 18±1 | | | | |
| [NeIII] | 4056 | | | | 26±2 | 43±4 | | | |
| H$\delta$ | 4198 | | | | 15±1 | | | | |
| H$\gamma$ | 4437 | | | 43±6 | 52±5 | 49±7 | | | |
| [OIII] | 4459 | | | 28±4 | 35±5 | 26±5 | | | |
| HeII | 4790 | | | | 35±8 | 30±5 | | | |
| H$\beta$ | 4970 | 100 | 100 | 100 | 100 | 100 | 100 | 100 | 100 |
| [OIII] | 5070 | 112±16 | 114±13 | 170±18 | 261±9 | 332±33 | 94±23 | 42±6 | 36±6 |
| [OIII] | 5121 | 376±56 | 354±36 | 533±26 | 931±29 | 740±73 | 269±31 | 123±21 | 36±8 |
| [OI] | 6440 | | | 17.7±1.6 | 29.2±3.0 | | | | |
| [OI] | 6511 | | | | 9.6±0.9 | | | | |
| [NII] | 6695 | 88±21 | 76±11 | 68±7 | 84±9 | 139±21 | 71± 16 | 46±8 | 36±7 |
| H$\alpha$ | 6708 | 319±34 | 295±28 | 266±14 | 249±12 | 477±49 | 240±28 | 283±43 | 287±46 |
| [NII] | 6727 | 285±39 | 211±22 | 189±11 | 230±11 | 431±45 | 247±29 | 126±20 | 133±21 |
| [SII] | 6863 | 84±5 | 76±5 | 51±4 | 45±3 | 118±14 | 72±9 | 52±12 | 48±8 |
| [SII] | 6881 | 49±8 | 52±4 | 43±3 | 55±4 | 82±11 | 42±6 | 33±8 | 31±8 |
| [ArIII] | 7296 | | | | 16±1 | | | | |
| F(H$\beta$)$^{(b)}$ | 4970 | 2.84±0.3 | 13.5±1.0 | 34.0±7.2 | 522.1±15.8 | 134.4±12.9 | 11.25±1.1 | 5.2±0.6 | 3.9±0.5 |

(a) Observed wavelength
(b) Flux of narrow component of H$\beta$ in units of $10^{-17}$erg s$^{-1}$ cm$^{-2}$

**Table 5.** Line intensities relative to I($H\beta_n$) and errors in PA=50$^O$

| Ion | $\lambda^{(a)}$ | Region 1 | Region 2 | Region 3 | Region 4 | Region 5 | Region 6 | Region 7 | Region 8 |
|---|---|---|---|---|---|---|---|---|---|
| [OII] | 3813 | 221±41 | 180±31 | 241±24 | 221±11 | 261±22 | 209±29 | 96±11 | 112±10 |
| [NeIII] | 3956 | | | 57±8 | 87±5 | 99±10 | 39±10 | | |
| HeI | 3977 | | | | 29±3 | 41±6 | | | |
| [NeIII] | 4056 | | | | 32±2 | 15±3 | | | |
| H$\delta$ | 4198 | | | | 26±2 | 28±3 | | | |
| H$\gamma$ | 4437 | | | 61±14 | 63±4 | 56±12 | 36±6 | 25±6 | 22±7 |
| [OIII] | 4459 | | | 18±2 | 32±5 | 30±5 | | | |
| HeII | 4790 | | | | 31±4 | 27±5 | | | |
| H$\beta$ | 4970 | 100 | 100 | 100 | 100 | 100 | 100 | | |
| [OIII] | 5070 | 10±2 | 37±6 | 116±13 | 284±15 | 274±11 | 99±13 | 10±6 | 10±6 |
| [OIII] | 5121 | 41 ± 7 | 96 ± 15 | 370 ± 34 | 888 ± 45 | 875 ± 78 | 294 ± 48 | 31 ± 6 | 31 ± 5 |
| [OI] | 6440 | | | | 32±5 | 52±7 | | | |
| [OI] | 6511 | | | | 11±2 | 17±3 | | | |
| [NII] | 6695 | 56±15 | 80±13 | 57±8 | 84±17 | 107±34 | 84±14 | 42±6 | 62±9 |
| H$\alpha$ | 6708 | 276±36 | 268±27 | 241±22 | 289±21 | 519±52 | 231±33 | 302±19 | 395±22 |
| [NII] | 6727 | 141±21 | 187±22 | 208±20 | 300±21 | 383±44 | 233±32 | 124±9 | 163±11 |
| [SII] | 6863 | 65±11 | 64±9 | 58±7 | 41±4 | 99±11 | 76±12 | 41±8 | 52±6 |
| [SII] | 6881 | 36±4 | 50±8 | 49±8 | 43±4 | 68±8 | 44±9 | 27±6 | 34±5 |
| [ArIII] | 7296 | | | | 14±2 | 20±3 | | | |
| F(H$\beta$)$^{(b)}$ | 4970 | 9.9±1.2 | 12.6±1.4 | 24.8±2.2 | 403.7±19.4 | 111.7±9.4 | 13.7±1.8 | 32.5±2.1 | 33.6±1.8 |

(a) Observed wavelength
(b) Flux of narrow component of H$\beta$ in units of $10^{-17}$erg s$^{-1}$cm$^{-2}$

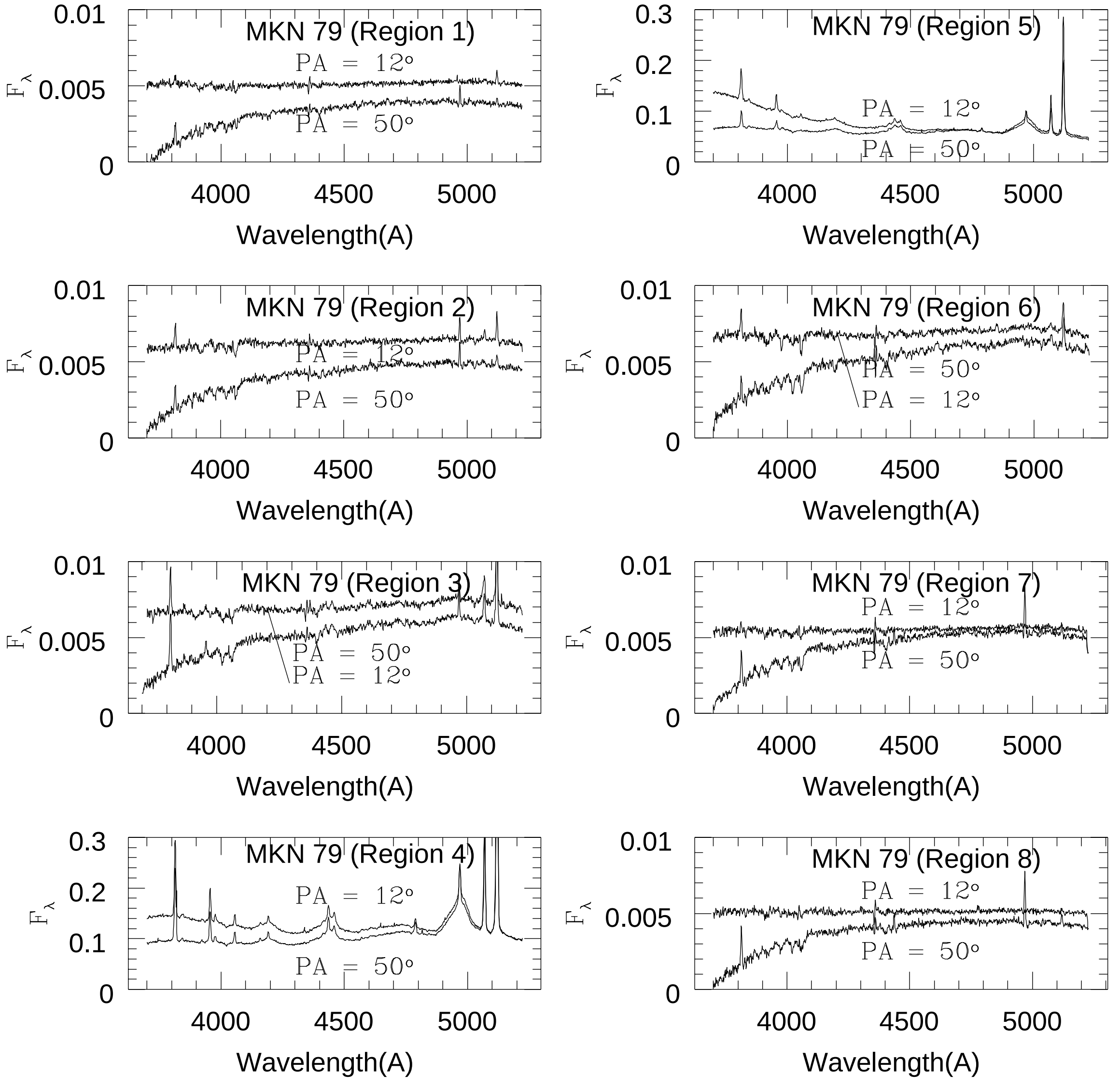

**Fig. 6.** The spectra of Regions 1 - 8 of Mkn79 The extraction windows correspond to 4 arcsec.The fluxes in units of $10^{-14}$erg $s^{-1}cm^{-2}Å^{-1}$

of the high ionization [OIII]$\lambda$5007 line to the low ionization [OII]$\lambda$3727, [NII]$\lambda$6583 and H$\beta$ lines in both position angles. The ratios of low ionization lines in both position angles are presented in Fig.8. In order to avoid involving widely separated lines, we formed the intensity ratios of low ionization lines in the blue and red wavelength ranges with H$\beta$ and H$\alpha$ respectively. The asymmetric structure in the high ionization line ratios seen in Fig. 7 are not seen in Fig. 8. Also, the high ionization line asymmetry is not seen at a comparable projected distance in PA=$50^o$. These results support our earlier statement regarding a higher level of gas excitation in the ENLR north of the nucleus in PA=$12^o$. The minima of the low ionization line ratios which are seen in the centre of the galaxy (Fig. 8), also indicate a higher level of gas ionization in the nucleus similar to that seen in NGC4151 (Robinson et al. 1994).

As poor seeing is the likely reason for the appearance of broad lines in regions 4 and 5, the HeII$\lambda$4686/H$\beta$ line intensity ratios in those regions could well describe conditions in the nucleus. The ratio of these lines in the nucleus and in regions 4 and 5 are similar, with an average value of 0.30±0.02. Assuming the existence of a relatively hard continuum in the nucleus of Mkn79, which causes the observed $I_{HeII}/I_{H\beta}$ ratio, we can estimate the spectral index of this powerlaw continuum using the observed line ratio. The ionizing photon energy for the HeII686 line ($h\nu \geq$54.6) is 4 times higher than that for H$\beta$ ($h\nu \geq$13.6). We can estimate, following Robinson et al. (1994), the ratio of ionizing photon luminosities $Q_{HII}$ and $Q_{H\beta}$ in the HeII$\lambda$4686 and H$\beta$ lines for a powerlaw continuum, $F_\nu \propto \nu^\alpha$, as

$$log(Q_{HeII}/Q_{H\beta}) = 0.6 \times \alpha,$$

where $\alpha$ is the spectral index.

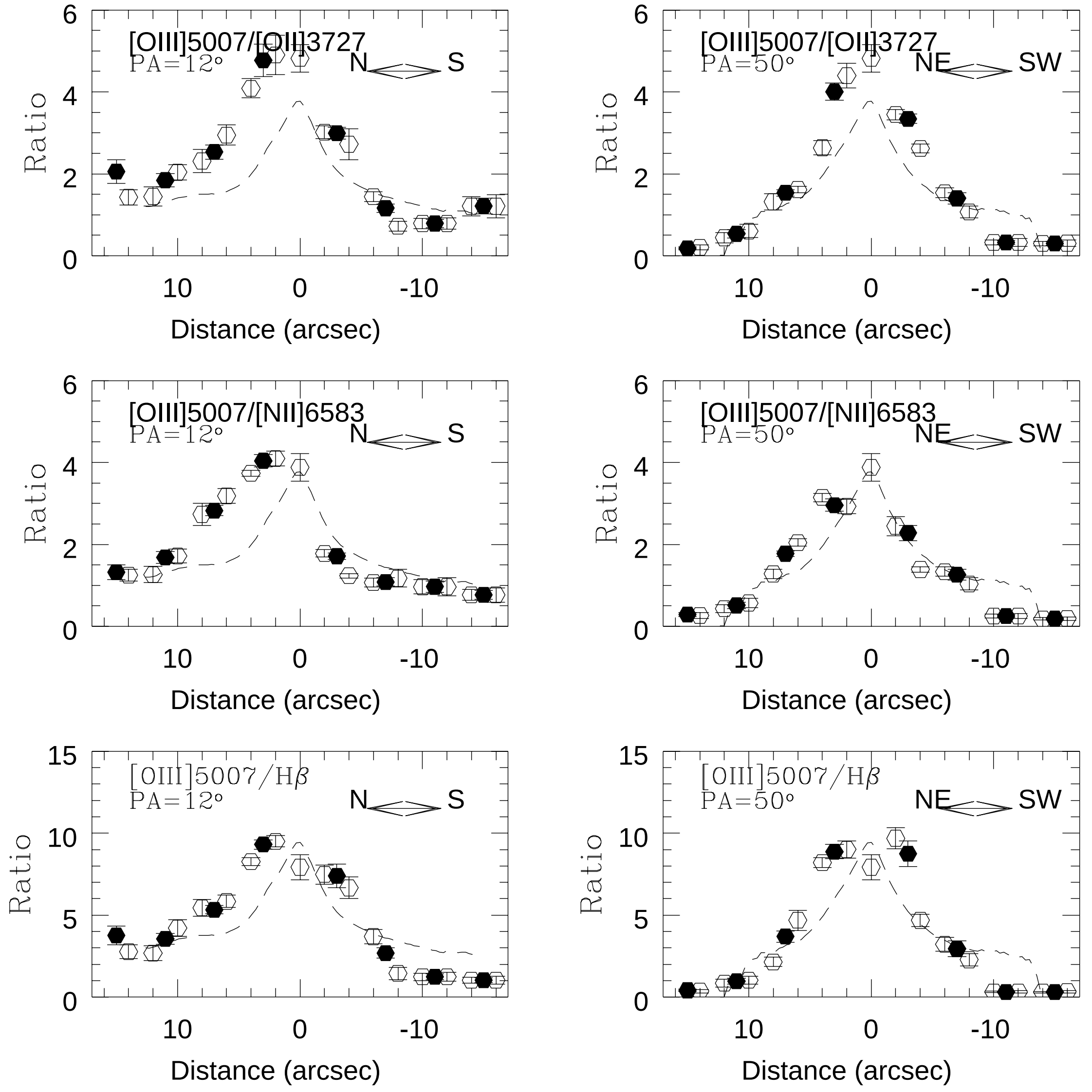

**Fig. 7.** Variation of the high ionization line intensity ratios with cross-section. Ratios obtained using 4 arcsec extraction windows are shown as filled circles. Open circles correspond to 2 arcsec windows. The log of the [OIII]$\lambda$5007 flux profiles in PA=12$^o$ and PA=50$^o$ are shown as dashed lines for comparison.

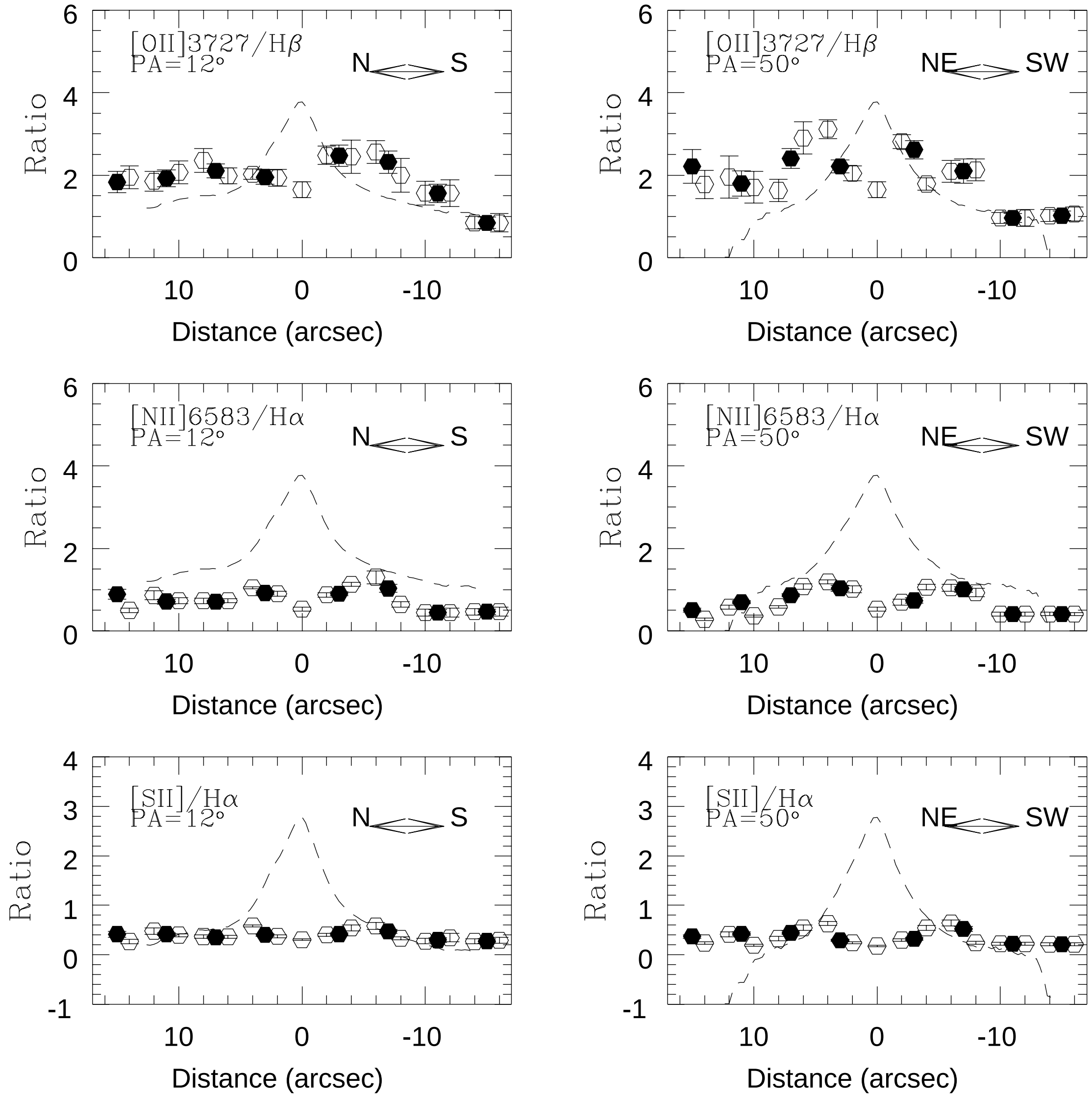

**Fig. 8.** Variation of the low ionization line intensity ratios with cross-section. Ratios obtained using 4 arcsec extraction windows are shown as filled circles. Open circles correspond to 2 arcsec windows. The log of the [OIII]$\lambda$5007 flux profiles in PA=12$^o$ and PA=50$^o$ are shown as dashed lines for comparison.

For the average observed line ratio, the required spectral index is $\alpha \leq -1.36\pm0.05$, in agreement with that derived from the low-resolution IUE spectra by Oke & Zimmerman (1979). The observed line ratio could be also consistent with a limiting blackbody ionizing continuum of temperature T$\leq1.25\times10^5$K (Binette et al. 1988).

*3.4. The ENLR Kinematics*

Using high resolution spectroscopy (spatial resolution of about 0.3 arcsec), Whittle et al. (1988) found double-component peak velocities for [OIII]$\lambda$5007 in PA=10$^o$ across the nucleus, with shifts relative to a systemic velocity of +100 km s$^{-1}$ and $-50$ km s$^{-1}$. The separation between the two velocity peaks is about 2-3 arcsec. Our data, shown in Fig. 9 using an extraction window of 1 arcsec, also show the changes in velocity, with a relative shift of 150 km s$^{-1}$ in the [OIII]$\lambda$ 5007 line which is nearly the same in both directions. Because our spectra have poorer spatial resolution compared to the spectra studied by Whittle et al. (1988), and also suffer worse seeing, we cannot confirm the validity of the double-peaked effect in PA=12$^o$. The velocity dispersion across the image is V$_d$= 200 $\pm$75 km s$^{-1}$ in PA=12$^o$ and is similar in PA=50$^o$. The line-of-sight systematic velocity in the nucleus is 6675$\pm$25 km s$^{-1}$, which is close to the value obtained by Whittle et al. (1988).

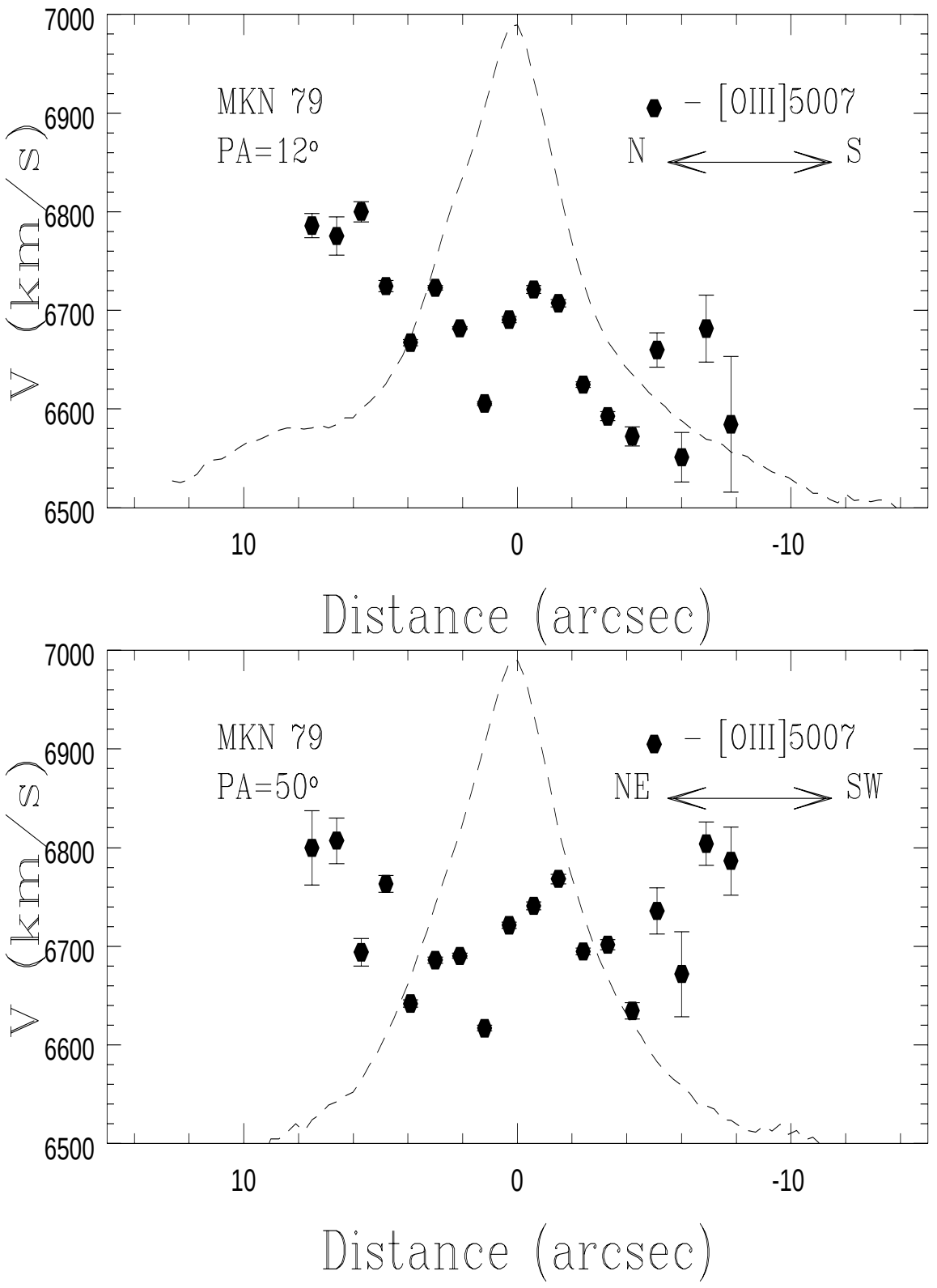

**Fig. 9.** Emission line velocities in PA=12$^o$ and PA=50$^o$.

## 4. Analysis of the results.

A major question in the investigation of the ENLR is understanding whether the emission lines arise by photoionization of the gas in the galactic disk by the central AGN ionizing continuum source, or whether there are other local sources for the gas ionization (young stars, extended ing between these various effects is to use line diagnostic diagrams (Baldwin et al.,1981; Veilleux & Osterbrock 1987; Robinson et al.,1987). These diagrams help give a clear understanding of the mechanism of gas heating and ionization, and can separate HII regions and planetary nebula from the ENLR and nuclei of AGN.

In order to investigate the physical conditions at different distances from the centre in Mkn79 we plot several diagnostic diagrams involving combinations of line ratios in Fig. 10. We also show the results of a series of photoionization models computed using the CLOUDY photoionization code (Ferland 1991) for plane parallel slabs of gas, of constant density N$_e$=100cm$^{-3}$, assuming solar abundances. Because the energy distribution of the ionizing continuum in the nucleus of Mkn79 is poorly known we made the calculations for both powerlaw and blackbody energy distributions. The powerlaw continuum has a spectral index $\alpha= -1.5$ in the range from 0.008 Ryd to 8000 Ryd. The blackbody continuum temperature is 1.3$\times$10$^5$K. Robinson et al. (1987) show that this continuum has a similar mean ionizing photon energy as a powerlaw continuum, with $\alpha$=$-$1.5 at approximately 35eV, and both produce line intensity ratios in agreement with the typical observed spectra of AGN. We calculated spectra for a wide range in ionization parameter $U$, where $U = F/(c \times N_e)$, and $F$ is the ionizing photon flux.

In the [NII]$\lambda$6583/H$\alpha$ and [SII]$\lambda$6717,31/H$\alpha$ versus [OIII]$\lambda$5007/H$\beta$ diagnostic diagrams, regions 1-6 in PA=12$^o$, including all regions north of the nucleus, and regions 3-6 in PA=50$^o$, are located in positions which are usually occupied by the narrow line regions of AGN (Veilleux & Osterbrock 1987). Again this indicates that the ENLR in Mkn79 has an elongation in position angle=12$^o$, but the ionizing cone from the centre could be quite wide, as in NGC4151 (Robinson et al. 1994) and includes several regions in PA=50$^o$. The regions south of the nucleus (12-7, 12-8) and also the regions associated with the strong hydrogen emission (50-1, 50-2, 50-7, 50-8) are located in positions corresponding to HII regions.

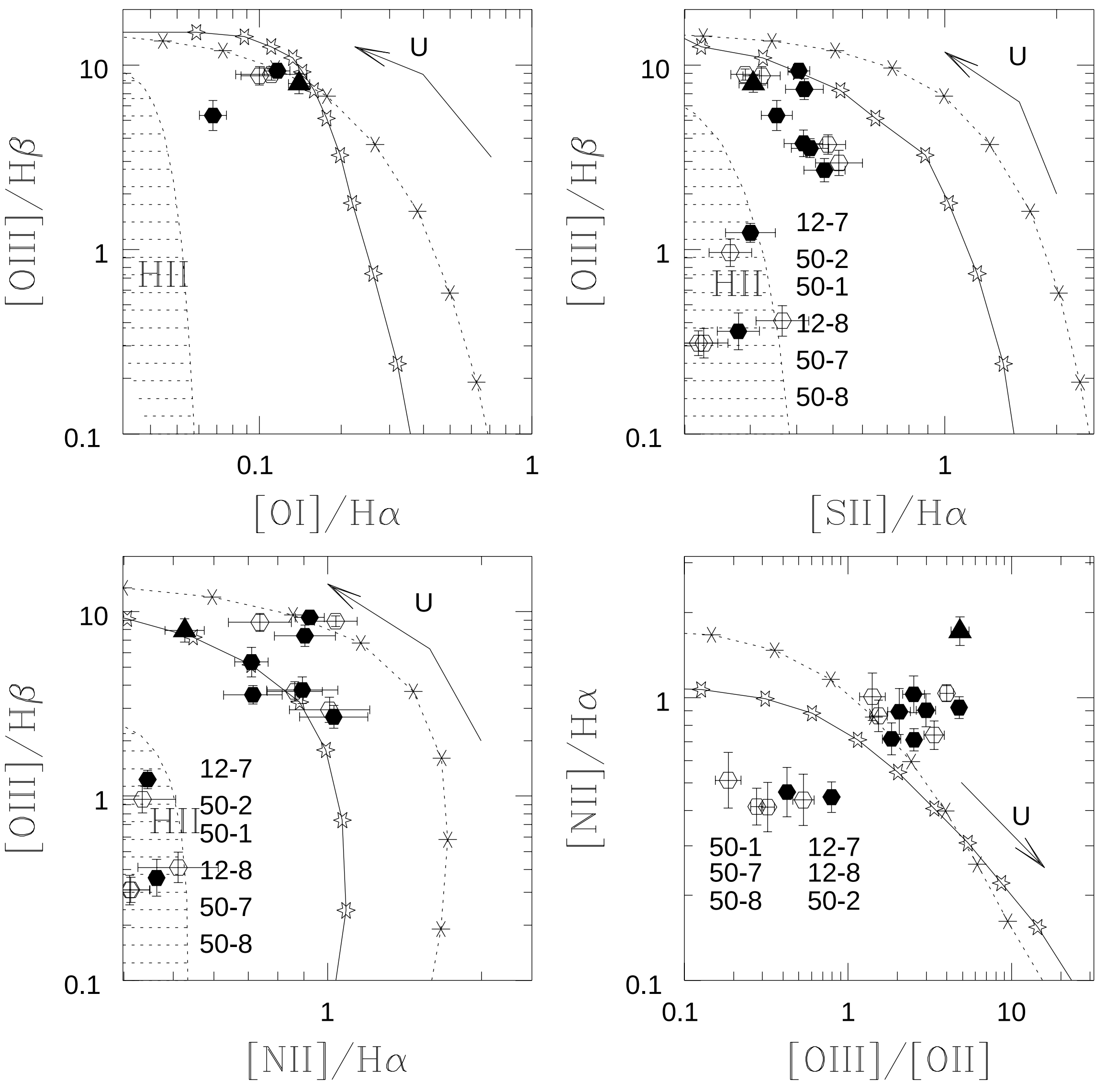

**Fig. 10.** Diagnostic diagrams for the emission regions of Mkn79 in both position angles. The shaded area indicates the regions occupied by extragalactic HII region-like objects (Veilleux & Osterbrock 1987). Filled circles indicate the regions in PA=$12^o$ and open circles show the regions in PA=$50^o$, with numbers corresponds to the regions listed in Table 2. The filled triangle shows the position of the nucleus. Photoionization model loci for the powerlaw ($\alpha=-1.5$) and blackbody continuum (T=$1.3\times10^5$) are plotted as solid and dashed lines respectively. The starred marks along these lines indicate successive increments of 0.25 in log$U$ starting at log$U$ = $-4.0$ except for the lower-right box which starts at $-3.75$.

In order to further constrain the ENLR phenomenon in Mkn79 we need to consider the ionizing photon budget by comparing the number of ionizing continuum photons available with the number required to produce the hydrogen line emission (Wilson et al., 1988). The number of ionising photons ($N_{H\beta}$) necessary to produce a H$\beta$ luminosity L$_{H\beta}$ is

$$N_{H\beta} = 2.1 \times 10^{52} (L_{H\beta}/10^{40} erg^{-1} s^{-1})\ photons\ s^{-1},$$

ditions. The H$\beta$ luminosity taken from our results is $3.8{\times}10^{41}$ erg s$^{-1}$ for a central region of $2 \times 1.5$ arcsec$^2$. The number of photons in a powerlaw ionizing continuum between $\nu_1$ and $\nu_2$ is

$$N_i = 4\pi R^2 C(\alpha h)^{-1}(\nu_1^{-\alpha} - \nu_2^{-\alpha})\ photons\ s^{-1}$$

where R = 132.86 Mpc is the distance to Mkn79 ($H_0$=50 km s$^{-1}$ Mpc$^{-1}$) and C is the constant in the powerlaw spectrum. We take $\nu_1 = 3.3{\times}10^{15}$ Hz and $\nu_2 = 4.8{\times}10^{17}$Hz and the spectral index $\alpha = -1.3$. The continuum luminosity of the central region within this range is $1.2{\times}10^{51}$ *photons s*$^{-1}$, assuming no interstellar reddening. However a colour excess of $\approx 0.22$ has been suggested for Mkn79 based on the $\lambda$2175 interstellar feature (Oke & Zimmerman 1979) and from line intensity ratios (Tsvetanov & Yancoulova 1989). Assuming this colour index, the continuum luminosity is $4.7{\times}10^{53}$ *photons s*$^{-1}$. Using the reddening corrected continuum luminosity and assuming a covering factor of 10% for the H$\beta$ emitting gas in the central region, the ratio $N_{H\beta}/Ni = 17$. This large ratio suggests that the ionizing continuum in Mkn79 is emitted anisotropically, being brighter towards the ENLR than in our direction. Alternatively, the continuum has a strong EUV/soft X-ray excess in addition to the powerlaw spectrum used above.

The radial ionization structure in PA=12$^o$ and PA=50$^o$ shows significant variation in the physical conditions along these directions. We can estimate the physical conditions at different distances from the nucleus, and in the nucleus itself, by the traditional method of using various line intensity ratios (Osterbrock 1989). The nuclear electron temperature can be derived from the line intensity ratio $I_{[OIII]\lambda4363}/I_{[OIII]\lambda\lambda5007,4959}$. The observed ratio corresponds to an electron temperature $T_e \geq 25000K \pm 5000K$, which is consistent with that obtained by Oke & Lauer (1979).

is shown in Fig.11. The observed NLR [SII]$\lambda$6717/[SII]$\lambda$6731 line ratio of $0.84 \pm 0.1$ corresponds to an electron density $N_e$ = 600 cm$^{-3}$ for $T_e$=25000K. The error-bars in the line ratio prohibit detailed analysis, but allow a rough estimation of the electron density in regions distant from the centre. The trend of the ratio is different north and south of the nucleus. There is a sharper increase south of the galaxy in both position angles between the NLR and the ENLR. The ratio appears constant, within the errors, south of the nucleus and corresponds to the low density limit, with $N_e < 80$ cm$^{-3}$. North of the nucleus the ratio indicates a smooth decrease in the electron density.

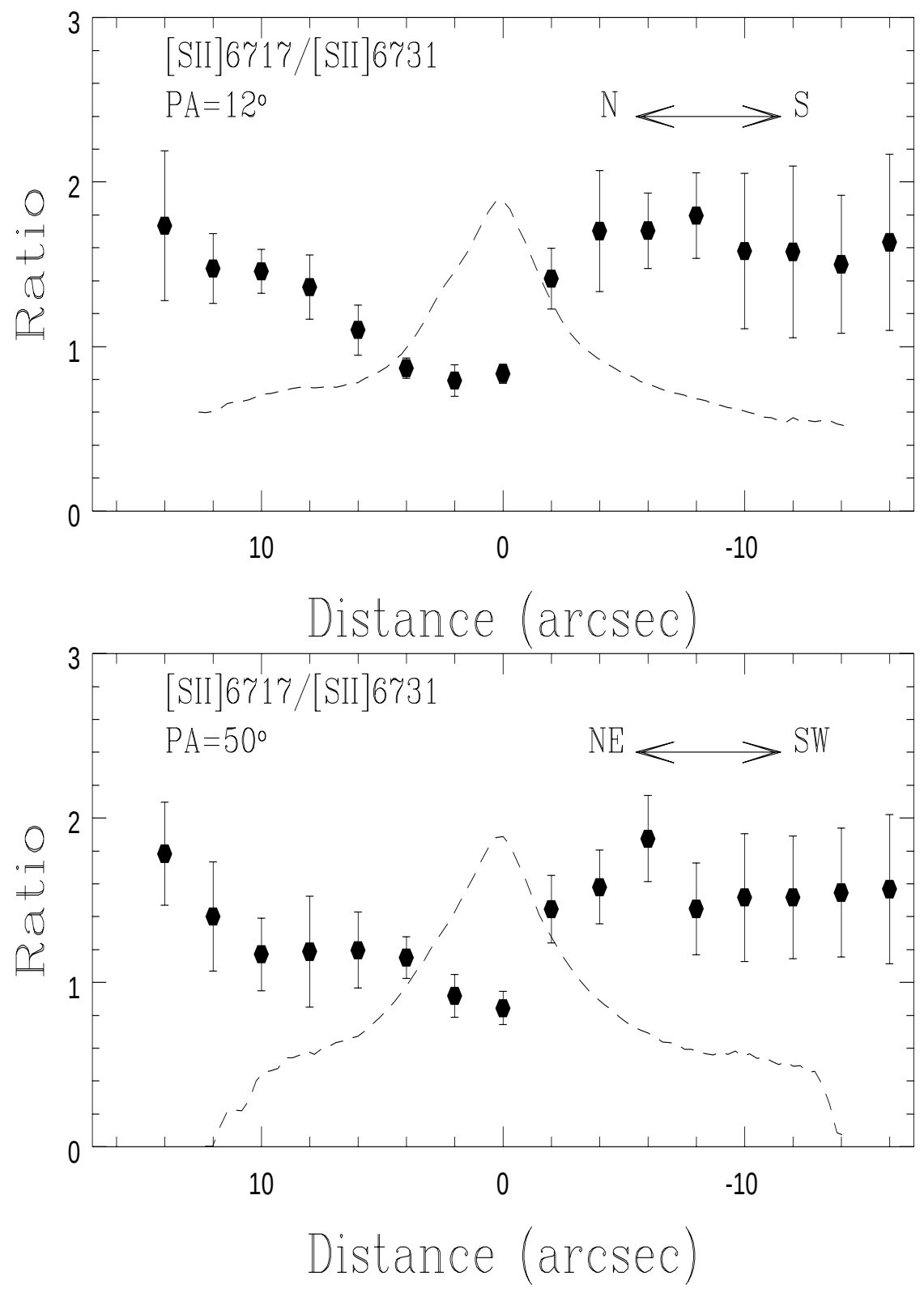

**Fig. 11.** Radial variation of the density-sensitive [SII]6717/[SII]6731 ratio in PA=12$^o$ and PA=50$^o$.

position angles, the electron density decreases by at least an order of magnitude, while the distance from the centre increases by a factor of 10. Thus the density appears to fall off by the inverse distance or more steeply, implying an ionization gradient across the northern ENLR. It should be noted that significant hydrogen line emission is also observed at a large distance from the centre in position angle PA=50$^o$. However, this emission is caused by the presence of HII regions in the host galaxy of Mkn79.

The source of the gas ionization in the ENLR of Mkn79 might be connected with the observed outflow of material from the nucleus of Mkn79 at PA=10$^o$ (Whittle et al. 1988) or with the radio structure at PA=12$^o$ ( Ulvestad & Wilson 1984). These and other possible explanations of our results, together with more detailed modelling of the ENLR in Mkn79 will be presented in a future paper.

## 5. Conclusions

We present deep long-slit spectra of Mkn79 at two position angles. The main results of our investigation of these spectra are as follows.

1. We find an extended narrow line region in Mkn79, particularly in PA=12$^o$.

2. The gas excitation is higher in the north in PA=12$^o$ compared to a comparable projected distance in PA=50$^o$. The average value of the Balmer decrement in the ENLR in both position angles is about 3.1, indicating that the ENLR in Mkn79 is not strongly reddened.

3. The Balmer decrement (I$(H\alpha)_b$/I$(H\beta)_b$) is 3.5 for the BLR, 3.8 for the NLR, and 3.1 for the ENLR. The electron temperature in the NLR derived from the line intensity ratio $I_{[OIII]\lambda4363}/I_{[OIII]\lambda\lambda5007,4959}$ is $T_e \geq 25000K \pm 5000K$, and the electron density estimated from the observed [SII]$\lambda$6717/[SII]$\lambda$6731 line ratio is $N_e = 600$ cm$^{-3}$.

[OIII]$\lambda$5007/H$\beta$ diagnostic diagrams the high ionization line regions are located in positions which usually occupied by narrow line regions of AGN (Veilleux & Osterbrock 1987). The distant regions south of the nucleus and also the regions associated with strong hydrogen line emission are located in positions usually occupied by HII regions.

5. The observed variation of the density-sensitive [SII]$\lambda$6717/[SII]$\lambda$6731 line ratio suggests a significant decrease in ENLR gas density with increasing distance north of the nucleus.

*Acknowledgements.* We are grateful to E.Terlevich for her kind help at the beginning of this project and R.Rutten for providing service observations. We want also to thank A. Robinson and referee F.Durret for very fruitful comments. L.N. wishes to thank the Royal Greenwich Observatory for their hospitality. This work has been partly supported by a Royal Astronomical Society grant.